\documentclass[3p,times,procedia]{elsarticle}
\usepackage{nupha_ecrc}

\volume{00}

\firstpage{1}

\journalname{Nuclear Physics A}

\runauth{}

\jid{nupha}

\jnltitlelogo{Nuclear Physics A}

\usepackage{amssymb}
\usepackage{amsmath}

\def\bp{{\boldsymbol p}}
\def\pperp{p_{\!\perp}}
\def\xperp{x_{\!\perp}}
\def\tg{\text{g}}
\def\tq{\text{q}}
\newcommand{\alt}{\protect\raisebox{-0.5ex}{$\:\stackrel{\textstyle <}{\sim}\:$}}
\newcommand{\agt}{\protect\raisebox{-0.5ex}{$\:\stackrel{\textstyle >}{\sim}\:$}}

\begin{document}

\begin{frontmatter}

\dochead{XXVIIth International Conference on Ultrarelativistic Nucleus-Nucleus Collisions\\ (Quark Matter 2018)}

\title{Nonequilibrium quark production in the expanding QCD plasma}


\author[label1]{Naoto Tanji}
\author[label2]{J\"{u}rgen Berges}

\address[label1]{European Centre for Theoretical Studies in Nuclear Physics and Related Areas (ECT*) and Fondazione Bruno Kessler, Strada delle Tabarelle 286, I-38123 Villazzano (TN), Italy}
\address[label2]{Institut f\"{u}r Theoretische Physik, Universit\"{a}t Heidelberg, Philosophenweg 16, 69120 Heidelberg, Germany}

\begin{abstract}
We present real-time lattice simulation results for nonequilibrium quark production from an over-occupied gluon plasma in longitudinally expanding geometry. 
The quark number density per unit transverse area and rapidity shows almost linear growth in time, and its growth rate appears to be consistent with a simple kinetic theory estimate involving only two-to-two scattering processes in small-angle approximation. We also find that quarks produced at early times satisfy a nonequilibrium scaling law.
\end{abstract}


\end{frontmatter}


\section{Introduction} \label{sec:intro}
In the high-energy limit of heavy-ion collisions, the system right after a collision is described as an over-occupied gluon plasma expanding in the beam direction. 
While the QCD coupling $\alpha_s $ is weak at high energies, the system is strongly correlated because the typical occupation number of gluons is inversely proportional to the coupling. 
First-principles-based descriptions of such nonequilibrium and strongly correlated systems are possible by means of classical-statistical gauge theory simulations \cite{Berges:2013fga}. 
In the pure gluonic sector, much progress has been made recently in theoretical descriptions of the early stage of heavy-ion collisions \cite{Mazeliauskas:QM18}. 

To find observable consequences of the early stage of heavy-ion collisions, the understanding of quark dynamics is crucial since they directly couple to electromagnetic probes. 
The pre-equilibrium quark dynamics has relevance also in investigations of the chiral magnetic effect \cite{Kharzeev:2015znc} as well as in the understanding of chemical equilibration between gluons and light quarks \cite{Gelis:2005pb}. 
In this contribution, we present the first results on $3+1$ dimensional real-time lattice simulations of nonequilibrium quark production in the longitudinally expanding QCD plasma \cite{Tanji:2017xiw}.

\section{Real-time lattice simulations} \label{sec:latt}
By a systematic weak-coupling expansion around strong gauge fields, one can derive evolution equations for classical-statistical gauge fields and dynamical quark fields based on the Schwinger--Keldysh path integral formalism \cite{Kasper:2014uaa}. Within this approximation, the over-occupied non-Abelian gauge fields obey the classical Yang--Mills equations
\begin{equation}
\left[ D_\mu \, \raisebox{-2pt}{,} \, F^{\mu\nu} \right] = J^\nu
\end{equation}
with fluctuating initial conditions. Here, $J^\nu$ stands for the color current induced by dynamical quarks corresponding to backreaction from quarks to the gauge fields. The quantum dynamics of quarks under the strong gauge fields is described by the Dirac equation for the quark mode functions:
\begin{equation}
\left( i\gamma^\mu D_\mu -m \right) \psi_n (x) = 0 \, ,
\end{equation}
where $n$ denotes quantum numbers of a quark. 
We have solved lattice-discretized versions of these equations in proper time $\tau=\sqrt{t^2-z^2}$ and rapidity $\eta=\text{atanh}(z/t)$ coordinates. 

To extract the information of the time evolution from the solutions of the evolution equations, we compute the momentum distribution functions for quarks $f_\tq (\tau ,\bp)$ and for gluons $f_\tg (\tau ,\bp )$ that are defined by the projection of the fields onto free modes. 
As an initial condition for the gauge sector, we assume the gauge field modes to have an initial distribution 
\begin{equation}
f_\tg (\tau_0, \pperp ,p_z) = \frac{1}{g^2} \Theta \left( Q_s -\sqrt{\pperp^2 +(\xi_0 p_z)^2} \right) \, ,
\end{equation}
where $Q_s$ is the characteristic momentum (saturation) scale and 
$\xi_0$ is the initial anisotropy parameter. The over-occupied gluon plasma characterized by such a distribution is expected to form at a time $\tau_0 \sim Q_s^{-1} \ln^2 \alpha_s^{-1}$ after the coherent initial gauge fields decay due to instabilities \cite{Romatschke:2006nk}. 
For the quark sector, the vacuum initial condition that corresponds to $f_\tq (\tau_0 ,\bp)=0$ is assumed. 

More details of the formulation can be found in Ref.~\cite{Tanji:2017xiw}. 

\begin{figure}[tb]
 \begin{center}
  \includegraphics[clip,width=7cm]{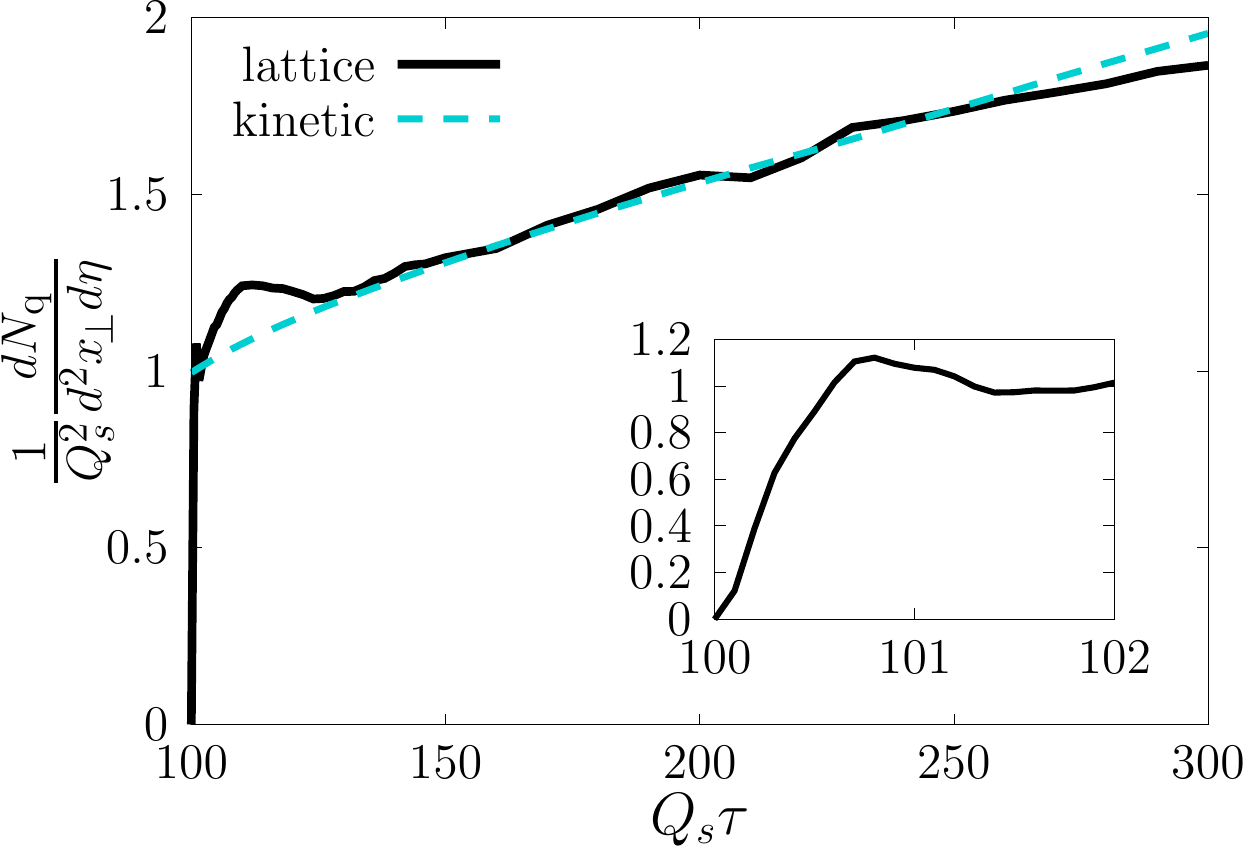} 
 \end{center}
 \vspace{-10pt}
\caption{Time evolution of the quark number density. The inset shows the early-time behavior.
The lattice result is compared to a simple kinetic theory estimate.}
\label{fig:qnum}
\end{figure}

\section{Numerical results} \label{sec:results}
We present numerical results for the real-time lattice QCD simulations with $N_c=2$ and $N_f=1$. The values of the coupling constant, the initial time and the anisotropy parameter are fixed to $g=10^{-2}$, $Q_s \tau_0 =100$ and $\xi_0 =2$. 
For results with large $N_f$ such that $N_f g^2$ is order one, we refer to Ref.~\cite{Tanji:2017xiw}. 

In Fig.~\ref{fig:qnum}, the total quark number density per unit transverse area and unit rapidity 
\begin{equation}
\frac{dN_\tq}{d^2 \xperp d\eta} = 4N_c N_f \, \tau \int \! \frac{d^3 p}{(2\pi)^3} \, f_\tq (\tau ,\bp )
\end{equation}
is plotted as a function of time $\tau$. As shown in the inset, it exhibits rapid increase at early times $Q_s \tau \alt 101$,  which can be interpreted as nonperturbative particle production from the initial quench. 
At later times $Q_s \tau \agt 130$, the total quark number density increases almost linearly in time. 
Remarkably, this linearly increasing behavior may be well explained by an effective kinetic theory description. 
In Fig.~\ref{fig:qnum}, a simple kinetic theory estimate that involves only two-to-two scattering processes with small-angle approximation is plotted for comparison with the lattice result. The production rate is well reproduced by the kinetic theory estimate. We emphasize that there is a priori no reason to expect the kinetic description to agree with the lattice results even in the order of magnitude accuracy, because the occupation number of gluons $\sim 1/g^2$ is nonperturbatively large. This observation indicates that the effective kinetic theory description may be more robust than one can expect based on perturbative power counting. 

\begin{figure}[tb]
 \begin{tabular}{cc}
 \begin{minipage}{0.48\hsize}
  \begin{center}
   \includegraphics[clip,width=7cm]{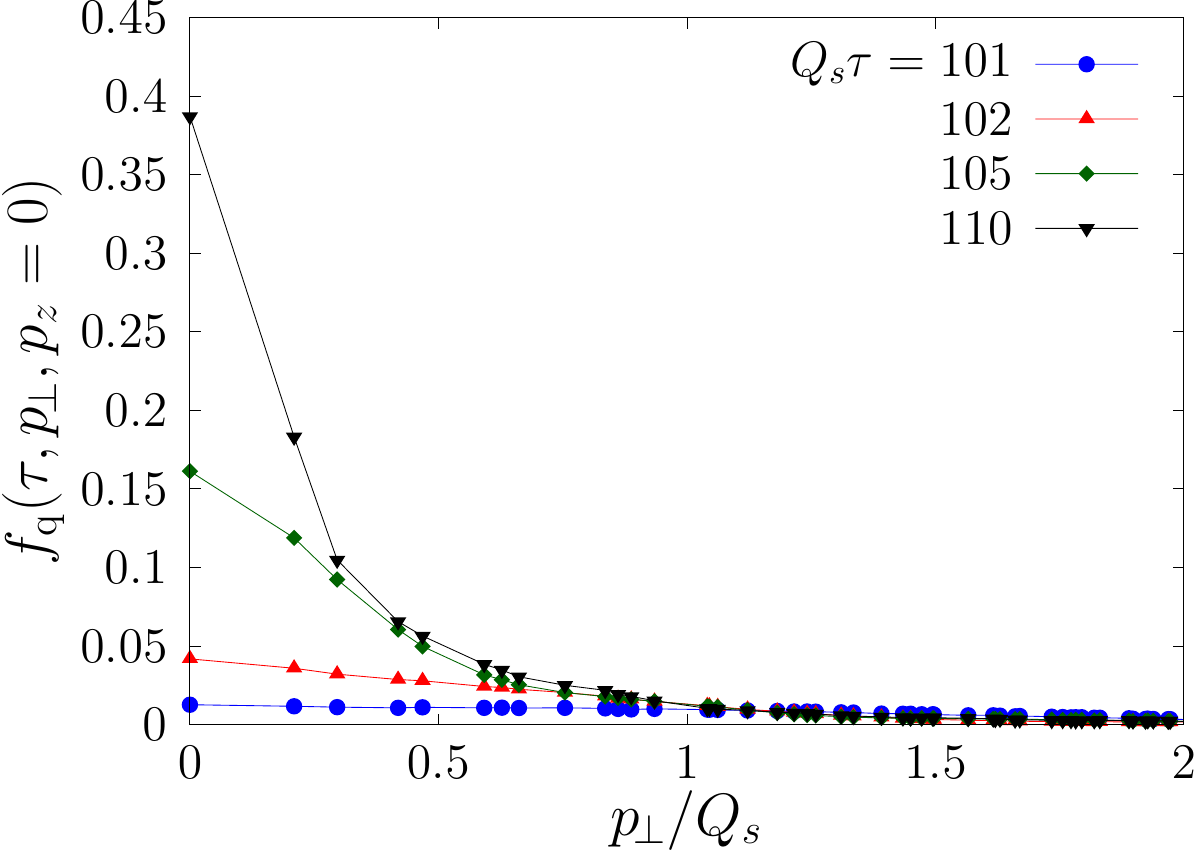} 
  \end{center}
 \end{minipage} &
 \begin{minipage}{0.48\hsize}
  \begin{center}
   \includegraphics[clip,width=7cm]{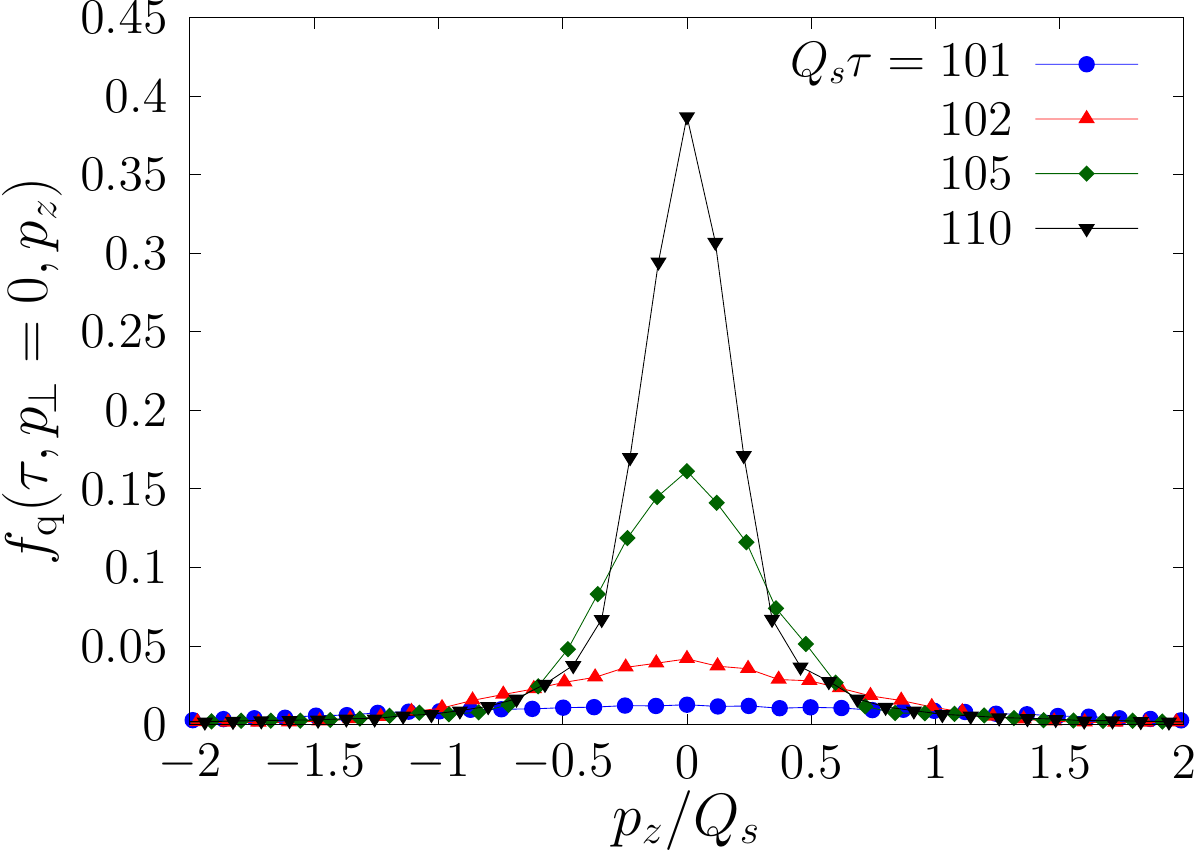} 
  \end{center}
 \end{minipage} 
 \end{tabular}
\caption{The quark distribution function at earlier times $Q_s \tau \leq 110$.
Left: The transverse momentum distribution at $p_z=0$. Right: The longitudinal momentum distribution at $\pperp =0$.}
\label{fig:fq1}
\end{figure}
\begin{figure}[tb]
 \begin{tabular}{cc}
 \begin{minipage}{0.48\hsize}
  \begin{center}
   \includegraphics[clip,width=7cm]{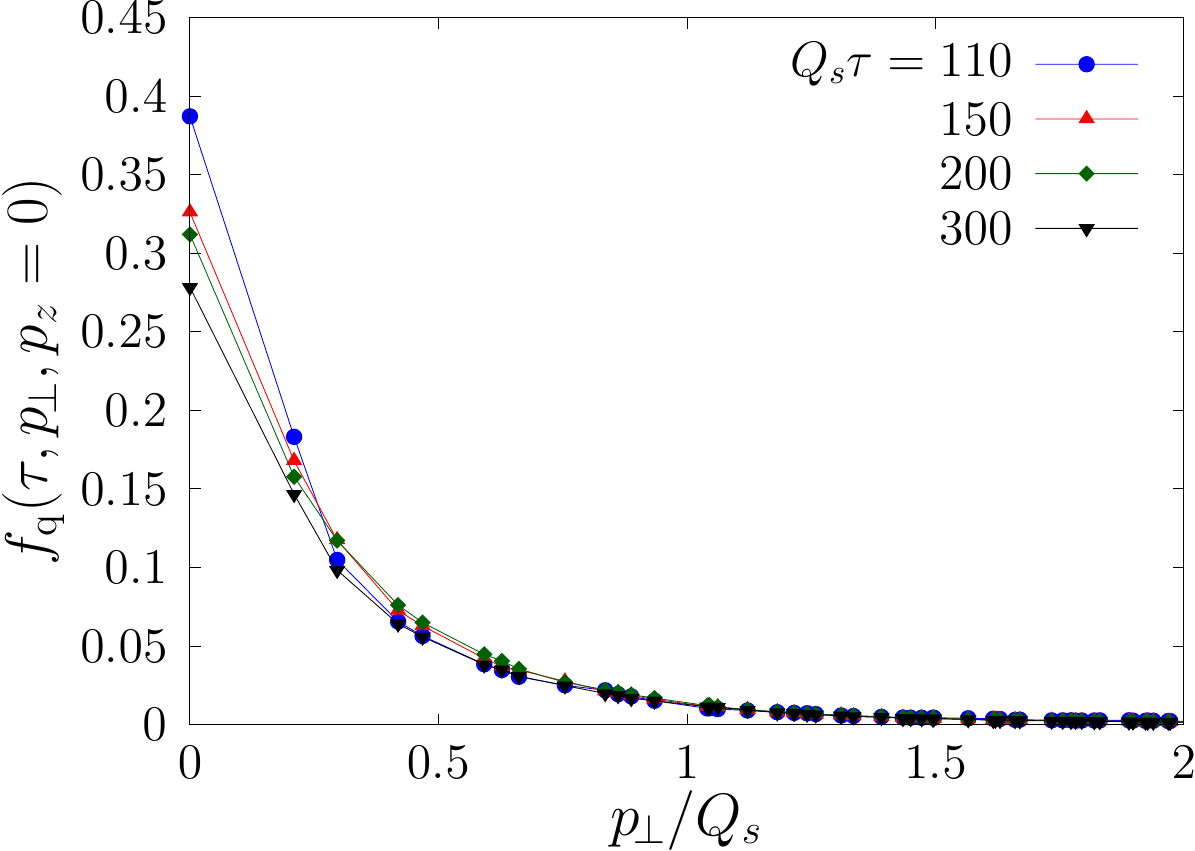} 
  \end{center}
 \end{minipage} &
 \begin{minipage}{0.48\hsize}
  \begin{center}
   \includegraphics[clip,width=7cm]{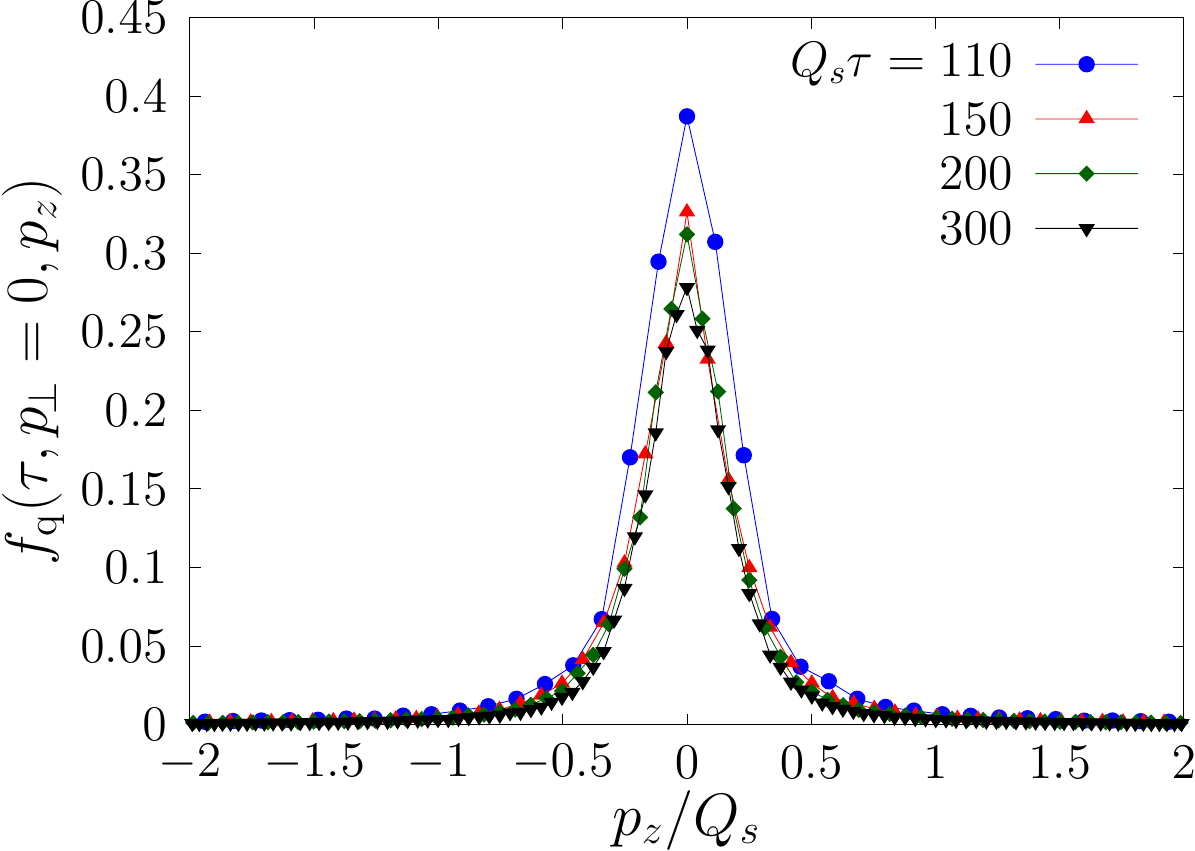} 
  \end{center}
 \end{minipage} 
 \end{tabular}
\caption{The quark distribution function at later times $Q_s \tau \geq 110$.
Left: The transverse momentum distribution at $p_z=0$. Right: The longitudinal momentum distribution at $\pperp =0$.}
\label{fig:fq2}
\end{figure}

Fig.~\ref{fig:fq1} shows the quark distribution function at earlier times $Q_s \tau \leq 110$. In this time range, an occupation number of order one is quickly developed. This result confirms that the quark production is seizable even at weak coupling,  as expected from the presence of highly occupied gluons. 

The evolution behavior at later times $Q_s \geq 110$ shown in Fig.~\ref{fig:fq2} looks quite different from that at earlier times. 
Remarkably, the width of the longitudinal momentum distribution becomes almost constant for $Q_s \tau \geq 150$. This behavior is in contrast to free-streaming, where the width of the longitudinal distribution shrinks in time as $p_z \sim 1/\tau$ due to the expansion of the system. 
The nonequilibrium steady state characterized by the constant width emerges because the effect of the momentum broadening caused by the particle production and scattering is balanced with the effect of the system expansion. 
This observation is consistent with the linearly increasing behavior of the total quark number density seen in Fig.~\ref{fig:qnum}. 

\begin{figure}[tb]
 \begin{tabular}{cc}
 \begin{minipage}{0.48\hsize}
  \begin{center}
   \includegraphics[clip,width=7cm]{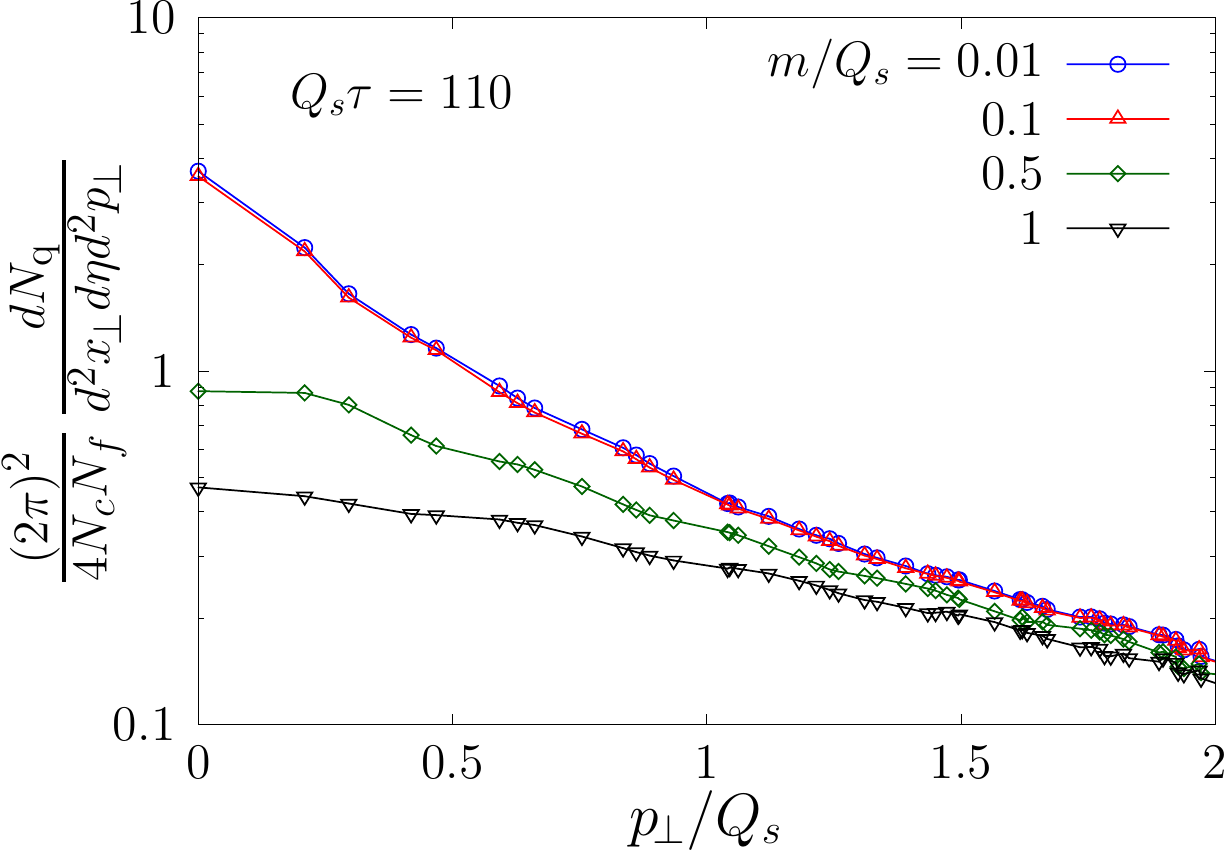} 
  \end{center}
 \end{minipage} &
 \begin{minipage}{0.48\hsize}
  \begin{center}
   \includegraphics[clip,width=7cm]{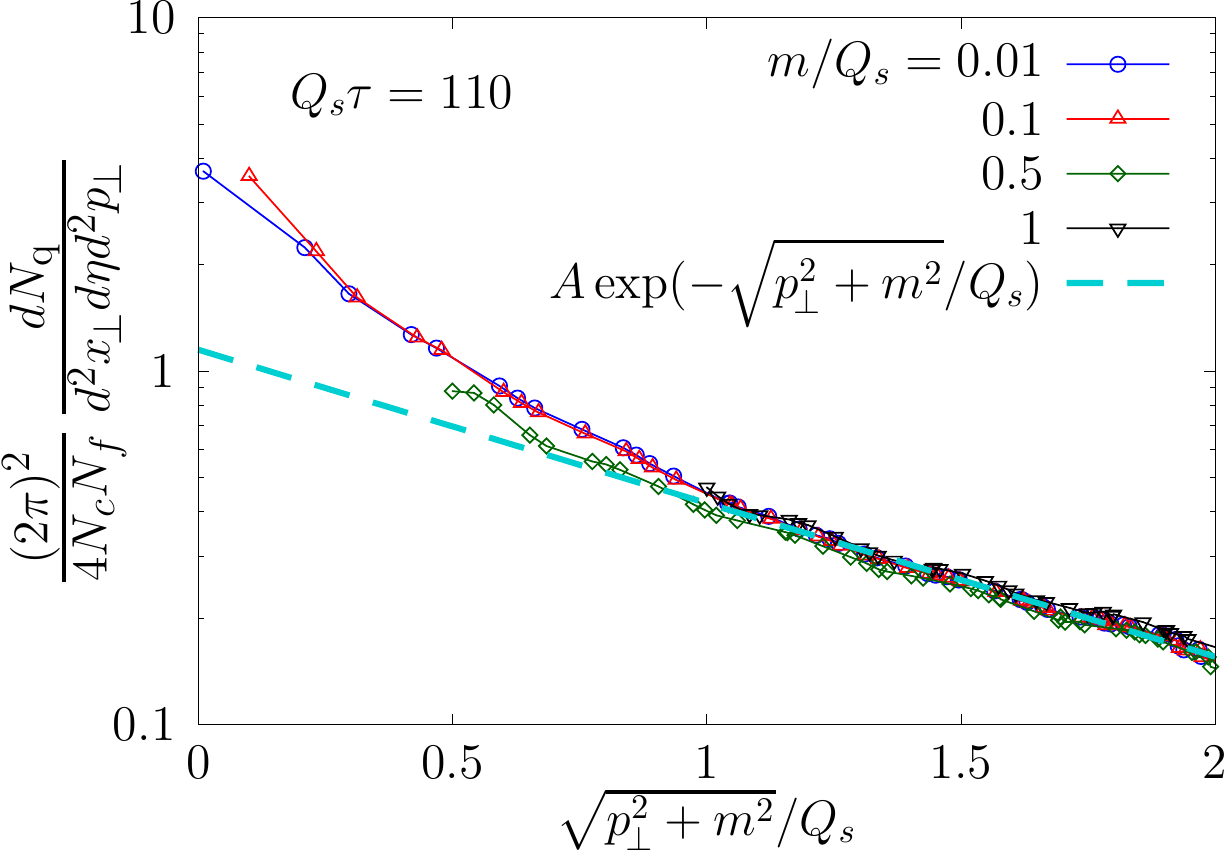} 
  \end{center}
 \end{minipage} 
 \end{tabular}
\caption{The integrated transverse spectrum for different quark masses at $Q_s \tau =110$ as a function of transverse momentum (left) and of transverse mass (right). In the right panel, an exponential function $A \exp(-\sqrt{\pperp^2 +m^2}/Q_s)$ is plotted as a dashed line with fit value $A=1.15$.}
\label{fig:mT}
\end{figure}

The quark mass dependence of the transverse spectra at time $Q_s \tau =110$ is depicted in Fig.~\ref{fig:mT}. The left panel shows the spectra as a function of $\pperp$. The production of heavy quarks are suppressed and the natural mass ordering can be seen. The spectra for $m/Q_s =0.01$ and 0.1 are almost degenerated. 
For $Q_s \simeq 1$ GeV, this result indicates that the production rate of the three light quark flavors in QCD are comparable. 
In the right panel, the spectra are plotted as a function of transverse mass $m_T =\sqrt{\pperp^2 +m^2}$. 
Interestingly, all data points for different masses lie on top of each other in the region $m_T \agt Q_s$. This means that the quark transverse spectrum in this momentum region is only a function of transverse mass. Furthermore, the shape of the overlapped region is consistent with an exponential function with an inverse slope $Q_s$. 
We note that this nonequilibrium scaling law can be seen only at earlier times when the nonperturbative rapid quark production is dominant. 
At later times, the transverse-mass scaling becomes less obvious. This is because the later-time quark production is well described by two-to-two scattering processes, and the kinetic processes in general do not satisfy the transverse-mass scaling. 

\section{Summary} \label{sec:summary}
We have investigated the nonequilibrium evolution of dynamical quarks and classical-statistical gauge fields in  longitudinally expanding geometry by using real-time lattice simulation techniques. The quark production from the over-occupied gluon plasma can be characterized by two temporal stages. In the earlier time regime, rapid and abundant quark production happens and the quark momentum distribution satisfies a nonequilibrium scaling law. In the later stage, the total quark number density shows an almost linear growth in time, which appears to be consistent with a simple kinetic estimate that includes only two-to-two scatterings.



\begin{thebibliography}{1}
\expandafter\ifx\csname url\endcsname\relax
  \def\url#1{\texttt{#1}}\fi
\expandafter\ifx\csname urlprefix\endcsname\relax\def\urlprefix{URL }\fi
\expandafter\ifx\csname href\endcsname\relax
  \def\href#1#2{#2} \def\path#1{#1}\fi

\bibitem{Berges:2013fga}
J.~Berges, K.~Boguslavski, S.~Schlichting, R.~Venugopalan, {Universal attractor
  in a highly occupied non-Abelian plasma}, Phys. Rev. D89~(11) (2014) 114007.
\newblock \href {http://arxiv.org/abs/1311.3005} {\path{arXiv:1311.3005}},
  \href {http://dx.doi.org/10.1103/PhysRevD.89.114007}
  {\path{doi:10.1103/PhysRevD.89.114007}}.

\bibitem{Mazeliauskas:QM18}
A.~Mazeliauskas, {these proceedings}.

\bibitem{Kharzeev:2015znc}
D.~E. Kharzeev, J.~Liao, S.~A. Voloshin, G.~Wang, {Chiral magnetic and vortical
  effects in high-energy nuclear collisions?A status report}, Prog. Part. Nucl.
  Phys. 88 (2016) 1--28.
\newblock \href {http://arxiv.org/abs/1511.04050} {\path{arXiv:1511.04050}},
  \href {http://dx.doi.org/10.1016/j.ppnp.2016.01.001}
  {\path{doi:10.1016/j.ppnp.2016.01.001}}.

\bibitem{Gelis:2005pb}
F.~Gelis, K.~Kajantie, T.~Lappi, {Chemical thermalization in relativistic heavy
  ion collisions}, Phys. Rev. Lett. 96 (2006) 032304.
\newblock \href {http://arxiv.org/abs/hep-ph/0508229}
  {\path{arXiv:hep-ph/0508229}}, \href
  {http://dx.doi.org/10.1103/PhysRevLett.96.032304}
  {\path{doi:10.1103/PhysRevLett.96.032304}}.

\bibitem{Tanji:2017xiw}
N.~Tanji, J.~Berges, {Nonequilibrium quark production in the expanding QCD
  plasma}, Phys. Rev. D97~(3) (2018) 034013.
\newblock \href {http://arxiv.org/abs/1711.03445} {\path{arXiv:1711.03445}},
  \href {http://dx.doi.org/10.1103/PhysRevD.97.034013}
  {\path{doi:10.1103/PhysRevD.97.034013}}.

\bibitem{Kasper:2014uaa}
V.~Kasper, F.~Hebenstreit, J.~Berges, {Fermion production from real-time
  lattice gauge theory in the classical-statistical regime}, Phys. Rev. D90~(2)
  (2014) 025016.
\newblock \href {http://arxiv.org/abs/1403.4849} {\path{arXiv:1403.4849}},
  \href {http://dx.doi.org/10.1103/PhysRevD.90.025016}
  {\path{doi:10.1103/PhysRevD.90.025016}}.

\bibitem{Romatschke:2006nk}
P.~Romatschke, R.~Venugopalan, {The Unstable Glasma}, Phys. Rev. D74 (2006)
  045011.
\newblock \href {http://arxiv.org/abs/0605045[hep-ph]}
  {\path{arXiv:0605045[hep-ph]}}, \href
  {http://dx.doi.org/10.1103/PhysRevD.74.045011}
  {\path{doi:10.1103/PhysRevD.74.045011}}.

\end{thebibliography}

\end{document}